\documentclass[prl,twocolumn,showpacs]{revtex4}

\usepackage{amssymb,amsmath}

\usepackage{graphicx}

\begin{document}
\title{Brazil-nut effect versus reverse Brazil-nut effect in a moderately dense granular fluid}
\author{Vicente Garz\'o}
\email{vicenteg@unex.es} \homepage{http://www.unex.es/eweb/fisteor/vicente/} \affiliation{Departamento de
F\'{\i}sica, Universidad de Extremadura, E-06071 Badajoz, Spain}

\begin{abstract}

A new segregation criterion based on the inelastic Enskog kinetic equation is derived
to show the transition between the Brazil-nut effect (BNE) and the reverse Brazil-nut
effect (RBNE) by varying the different parameters of the system. In contrast to
previous theoretical attempts the approach is not limited to the near-elastic case,
takes into account the influence of both thermal gradients and gravity and applies for
moderate densities. The form of the phase-diagrams for the BNE/RBNE transition depends
sensitively on the value of gravity relative to the thermal gradient, so that it is
possible to switch between both states for given values of the mass and size ratios,
the coefficients of restitution and the solid volume fraction. In particular, the
influence of collisional dissipation on segregation becomes more important when the
thermal gradient dominates over gravity than in the opposite limit. The present
analysis extends previous results derived in the dilute limit case and is consistent
with the findings of some recent experimental results.
\end{abstract}

\draft 
\date{\today}
\maketitle

Segregation and mixing of dissimilar grains in agitated granular binary mixtures is one
of the most important problems in granular matter both from a fundamental and a
practical point of view. In some cases it is a desired and useful effect to separate
particles of different types, while in other processes it is undesired and can be
difficult to control. Usually, the larger intruder particles tend to climb to the top
of the sample against gravity (Brazil-nut effect, BNE), but under certain conditions
they can also accumulate at the bottom (reverse Brazil-nut effect, RBNE). However,
although there is an extensive observational evidence of these phenomena, much less is
known on the physical mechanisms involved in this problem \cite{K04}. Among the
different competing mechanisms proposed to explain the transition BNE$\Leftrightarrow$
RBNE \cite{DRC93,HQL01,BEKR03,SUKSS06}, thermal (Soret) diffusion becomes the most
relevant one at large shaking amplitude where the system resembles a granular gas and
kinetic theory becomes useful to study segregation. Some previous theoretical attempts
have been reported in the literature analyzing thermal diffusion effects on segregation
in bi-disperse granular systems. Nevertheless, these early contributions have been
restricted to elastic \cite{JY02} and quasielastic particles \cite{TAH03}, have
considered thermalized systems (and so the segregation dynamics of intruders is only
driven by the gravitational force) \cite{JY02,TAH03}, and/or have been limited to
dilute gases \cite{BRM05,G06,SGNT06}. The main goal of this paper is to propose a
theory based on a recent solution of the inelastic Enskog equation \cite{GDH07} that
covers some of the aspects not accounted for in previous works. The theory subsumes all
previous analyses \cite{JY02,TAH03,BRM05,G06}, which are recovered in the appropriate
limits. Furthermore, the theoretical predictions are in qualitative agreement with some
molecular dynamics (MD) simulation results \cite{BRM05,SUKSS06,GDH05} and are also
consistent with previous experimental works \cite{BEKR03,SUKSS06}.
\begin{figure}
\includegraphics[width=0.8 \columnwidth,angle=0]{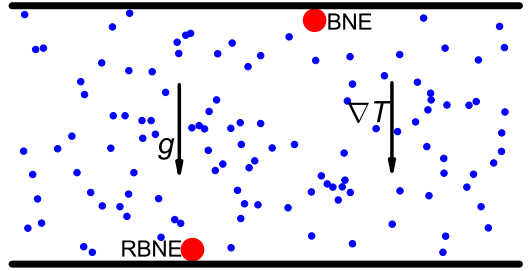}
\caption{(Color online) Sketch of the problem studied here. The small circles represent
the particles of the granular fluid while the large circles are the intruders. The BNE
(RBNE) effect corresponds to the situation in which the intruder rises (falls) to the
top (bottom) plate. \label{fig1}}
\end{figure}

We consider a binary mixture of inelastic hard disks ($d=2$) or spheres ($d=3$) where
one of the components (of mass $m_0$ and diameter $\sigma_0$) is present in tracer
concentration. In this limit case, one can assume that (i) the state of the dense gas
(excess component of mass $m$ and diameter $\sigma<\sigma_0$) is not affected by the
presence of tracer particles and, (ii) one can also neglect collisions among tracer
particles in their kinetic equation. This is formally equivalent to study an intruder
in a dense granular gas, and this will be the terminology used here. Collisions among
gas--gas and intruder-gas particles are inelastic and are characterized by two
independent (constant) coefficients of normal restitution $\alpha$ and $\alpha_0$,
respectively. The system is in presence of the gravitational field ${\bf g}=-g
\hat{{\bf e}}_z$, where $g$ is a positive constant and $\hat{{\bf e}}_z$ is the unit
vector in the positive direction of the $z$ axis. To fluidize the mixture and reach a
steady state, in most of the experiments energy is injected into the system through
vibrating horizontal walls. Here, in order to avoid the use of vibrating boundary
conditions, particles are assumed to be heated by a stochastic-driving force which
mimics a thermal bath. Although previous experiments \cite{FM02} have shown a less
significant dependence of the temperature ratio $T_0/T$ on inelasticity than the one
obtained in driven steady states \cite{DHGD02}, the results derived in Ref.\ \cite{G06}
for $T_0/T$ from this stochastic driving method compare quite well with MD simulations
of agitated mixtures \cite{SUKSS06}. This agreement suggests that this driving method
can be seen as a plausible approximation for comparison with experimental results. A
sketch of the geometry of the problem is given in Fig.\ \ref{fig1}.

The thermal (Soret) diffusion factor $\Lambda$ is defined at the steady state with zero
flow velocity and gradients only along the vertical direction ($z$ axis). Under these
conditions, the factor $\Lambda$ is defined by \cite{G06}
\begin{equation}
\label{1} -\Lambda \partial_z \ln T=\partial_z\ln \left(\frac{n_0}{n}\right),
\end{equation}
where $n_0$ and $n$ are the number densities of the intruder and the fluid particles,
respectively. Let us assume that gravity and thermal gradient point in parallel
directions (i.e., the bottom plate is hotter than the top plate, $\partial_z\ln T<0$)
(see Fig.\ \ref{fig1}). Obviously, when $\Lambda>0$, the intruder rises with respect to
the fluid particles (BNE, i.e., $\partial_z\ln (n_0/n)>0$) while if $\Lambda<0$, the
intruder falls with respect to the fluid particles (RBNE, i.e., $\partial_z\ln
(n_0/n)<0$).

\begin{figure}
\includegraphics[width=0.7 \columnwidth,angle=0]{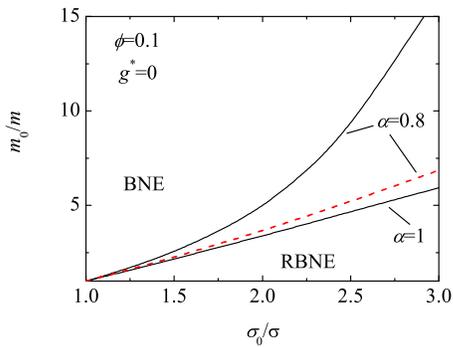}
\caption{(Color online) Phase diagram for BNE/RBNE for $\phi=0.1$ in the absence of
gravity for two values of the (common) coefficient of restitution $\alpha$. Points
above the curve correspond to $\Lambda>0$ (BNE) while points below the curve correspond
to $\Lambda<0$ (RBNE). The dashed line is the result obtained for $\alpha=0.8$ assuming
energy equipartition ($T_0=T$). \label{fig2}}
\end{figure}
\begin{figure}
\includegraphics[width=0.7 \columnwidth,angle=0]{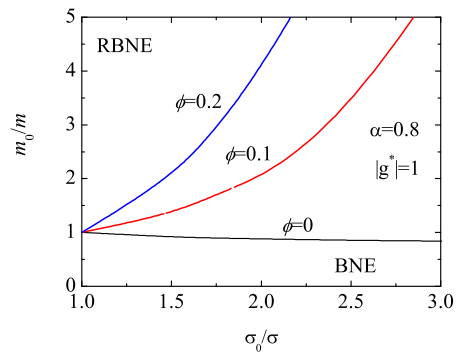}
\caption{(Color online) Phase diagram for BNE/RBNE for $\alpha=0.8$, $|g^*|=1$ and
three different values of the solid volume fraction $\phi$.  \label{fig3}}
\end{figure}
In order to determine the dependence of the coefficient $\Lambda$ on the parameters of
the system, we focus our attention on an {\em inhomogeneous} steady state with
gradients along the $z$ direction. Since the flow velocity vanishes, the mass flux of
the intruder $j_z=0$. In addition, the momentum balance equation yields
\begin{equation}
\label{2} \frac{\partial p}{\partial z}=\frac{\partial p}{\partial
T}\partial_zT+\frac{\partial p}{\partial
n}\partial_z n=-\rho g,
\end{equation}
where $p$ is the pressure and $\rho=m n$ is the mass density of the fluid particles.
Upon writing Eq.\ (\ref{2}), we have taken into account that $p$ depends on $z$ through
its dependence on $n$ and $T$ \cite{GDH07}. To first order in the spatial gradients
(Navier-Stokes description), the constitutive equation for the mass flux of the
intruder is \cite{GDH07}
\begin{equation}
\label{3} j_{z}=-\frac{m_0^2}{\rho}D_{0}
\partial_z n_0-\frac{m_0 m}{\rho}D\partial_z n-
\frac{\rho}{T}D^T\partial_zT,
\end{equation}
where $D_{0}$, $D$, and $D^T$ are the relevant transport coefficients. Expressions for
the pressure $p$ and the transport coefficients $D_{0}$, $D$, and $D^{T}$ have been
recently obtained in the undriven case by solving the Enskog kinetic equation by means
of the Chapman-Enskog method in the first Sonine approximation \cite{GDH07}. The
extension of these results to the driven case is straightforward. The condition
$j_{z}=0$ along with the balance equation (\ref{2}) allow one to get the thermal
diffusion factor $\Lambda$ in terms of the parameters of the mixture as
\begin{equation}
\label{4} \Lambda=\frac{\beta D^{T*}-(p^*+g^*)(D_{0}^*+D^*)}{\beta D_{0}^*}.
\end{equation}
\begin{figure}
\includegraphics[width=0.7 \columnwidth,angle=0]{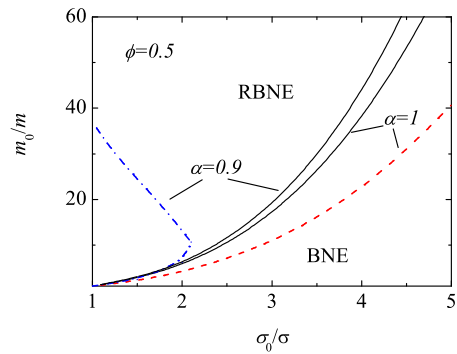}
\caption{(Color online) Phase diagram for BNE/RBNE for $\phi=0.5$ in the absence of
thermal gradient ($|g^*|\to \infty$) for two values of $\alpha$. The dashed and
dashed--dotted lines refer to the results obtained by Jenkins and Yoon \cite{JY02} and
Trujillo {\em et al.} \cite{TAH03}, respectively. \label{fig4}}
\end{figure}
Here, $\beta\equiv p^*+\phi \partial_\phi p^*$, $g^*=\rho g/n\partial_zT<0$ is a
dimensionless parameter measuring the gravity relative to the thermal gradient, and
$p^*=p/nT=1+2^{d-2}\chi\phi(1+\alpha)$. Moreover, $\chi(\phi)$ is the pair correlation
function for the granular gas, $\phi=[\pi^{d/2}/2^{d-1}d\Gamma(d/2)] n\sigma^d$ is the
solid volume fraction and the reduced transport coefficients are explicitly given by
\begin{equation}
\label{5} D_{0}^*=\frac{\gamma}{\nu_D},
\end{equation}
\begin{equation}
\label{6} D^{T*}=-\frac{M}{\nu_D}\left(p^*-\frac{\gamma}{M}\right)+
\frac{(1+\omega)^d}{2\nu_D}\frac{M}{1+M} \chi _{0}\phi(1+\alpha _{0}),
\end{equation}
\begin{equation}
\label{7} D^{*}=-\frac{M}{\nu_D}\beta+\frac{1}{2\nu_D}
\frac{\gamma+M}{1+M}\frac{\phi}{T}\left(\frac{\partial\mu_0}{\partial
\phi}\right)_{T,n_0}(1+\alpha_{0}).
\end{equation}
Here, $\gamma\equiv T_0/T$ is the temperature ratio, $M\equiv m_0/m$ is the mass ratio,
$\omega\equiv \sigma_0/\sigma$ is the size ratio and
\begin{equation}
\label{7.1} \nu_D=\frac{\sqrt{2}\pi^{(d-1)/2}}{d\Gamma\left(\frac{d}{2}\right)}\chi_0
\sqrt{\frac{\gamma+M}{M(1+M)}}(1+\alpha_0).
\end{equation}
In addition, $\chi_{0}$ is the intruder-gas pair correlation function and $\mu_0$ is
the chemical potential of the intruder. When the granular gas is driven by means of a
stochastic thermostat, the temperature ratio $\gamma$ is determined \cite{G06} from the
requirement $\gamma \zeta_0=M\zeta$ where the expressions for the cooling rates
$\zeta_0$ and $\zeta$ in the local equilibrium approximation can be found in Ref.\
\cite{GDH07}.

The condition $\Lambda=0$ provides the segregation criterion for  the transition BNE
$\Leftrightarrow$ RBNE. Since $\beta$ and $D_{0}^*$ are both positive, then according
to (\ref{4}), $\text{sgn} (\Lambda)=\text{sgn}(\beta D^{T*}-(p^*+g^*)(D_{0}^*+D^*))$.
Consequently, taking into account Eqs.\ (\ref{5})--(\ref{7}), the segregation criterion
can be written as
\begin{eqnarray}
\label{11}& & g^*(\gamma-M \beta)-\gamma \phi \frac{\partial p^*}{\partial
\phi}+\frac{(1+\omega)^d}{2}\frac{M}{1+M}\chi
_{0}\phi(1+\alpha_{0})\nonumber\\
& & \times \left[\left(p^*+g^*\right)\frac{M+\gamma}{M}\Delta-\beta\right]=0,
\end{eqnarray}
where $\Delta\equiv [(1+\omega)^{-d}/T\chi_{0}](\partial_\phi \mu_0)_{T,n_0}$. Equation
(\ref{11}) gives the phase-diagram for the BNE/RBNE transition due to Soret diffusion
of an intruder in a moderately dense granular gas. This is the most relevant result of
this paper. The parameter space of the problem is six-fold: the dimensionless gravity
$g^*$, the mass ratio $m_0/m$, the ratio of diameters $\sigma_0/\sigma$, the solid
volume fraction $\phi$, and the coefficients of restitution $\alpha$ and $\alpha_{0}$.
The influence of density on segregation is accounted for by the second and third terms
in Eq.\ (\ref{11}). As expected, when $m_0=m$, $\sigma_0=\sigma$ and $\alpha=\alpha_0$,
the system (intruder plus gas) is monodisperse and the two species do not segregate.
This is consistent with Eq.\ (\ref{11}) since in this limit case $D^{T*}=D_0^*+D^*=0$
and so, the condition (\ref{11}) holds for any value of $\alpha$ and $\phi$. In the
dilute limit case ($\phi\to 0$), $\beta=p^*=1$ so that Eq.\ (\ref{11}) reads
\begin{equation}
\label{12}g^*(\gamma-M)=0.
\end{equation}
Note that in the absence of gravity ($g^*=0$), Eq.\ (\ref{12}) applies for any value of
the parameters of the system and so the intruder does not segregate in a dilute gas.
When $g^* \neq 0$, the solution to (\ref{12}) is $\gamma=M$. This result agrees with
the recent segregation condition derived from the Boltzmann equation \cite{G06,BRM05}.

In general, segregation is driven and sustained by both gravity and temperature
gradients. The combined effect of both parameters on Soret diffusion is through the
reduced parameter $g^*$. This parameter measures the competition between these two
mechanisms on segregation. In previous theoretical studies on dense
gases\cite{JY02,TAH03}, the temperature was assumed to be uniform in the bulk of the
system ($\partial_zT=0$) so that segregation of intruder was essentially driven by
gravity. This is quite an interesting limit since most of the experiments
\cite{HQL01,BEKR03,SBKR05} have been carried out under these conditions. In this limit
($|g^*|\to \infty$), Eq.\ (\ref{11}) becomes
\begin{equation}
\label{13} \frac{1+\frac{(1+\omega)^d}{2}\chi
_{0}\phi(1+\alpha_{0})\frac{\gamma+M}{1+M}\frac{\Delta}{\gamma}}
{1+2^{d-2}\chi\phi(1+\alpha)\left[1+\phi\partial_\phi\ln
(\phi\chi)\right]}\frac{T_0}{T}-\frac{m_0}{m}=0,
\end{equation}
while the segregation criterion found independently by Jenkins and Yoon \cite{JY02} (for an elastic system) and
by Trujillo {\em et al.} \cite{TAH03} is
\begin{equation}
\label{14}\frac{1+\frac{(1+\omega)^d}{2}\chi
_{0}\phi}{1+2^{d-1}\chi\phi}\frac{T_0}{T}-\frac{m_0}{m}=0.
\end{equation}
Equation\ (\ref{13}) reduces to Eq.\ (\ref{14}) when one (i) neglects the dependence on
inelasticity and assumes equipartition in certain terms, (ii) takes the approximation
$\Delta =1$ (which only applies for a dilute gas of mechanically equivalent particles),
and (iii) neglects high density corrections (last term in the denominator of
(\ref{13})). Thus, even in the particular limit $|g^*|\to \infty$, the criterion
(\ref{13}) is much more general than the one previously derived \cite{TAH03} since it
covers the complete range of the parameter space of the problem.

Henceforth and for the sake of concreteness, we assume a three-dimensional system with
$\alpha=\alpha_0$. In this case \cite{MCSL71}, $\chi=(1-\frac{1}{2}\phi)/(1-\phi)^3$
and
\begin{equation}
\label{15} \chi_{0}=\frac{1}{1-\phi}+3\frac{\omega}{1+\omega}\frac{\phi}{(1-\phi)^2}+2
\frac{\omega^2}{(1+\omega)^2}\frac{\phi^2}{(1-\phi)^3}.
\end{equation}
The expression for the chemical potential consistent with the above approximations can
be found in Ref.\ \cite{RG73}. A phase diagram delineating the regimes between BNE and
RBNE is shown in Fig.\ \ref{fig2} for $\phi=0.1$, $g^*=0$ and two values of the
(common) coefficient of restitution $\alpha$. We observe that, in the absence of
gravity, the main effect of dissipation is to reduce the size of the BNE. It is also
apparent that the RBNE is dominant at both small mass ratio and/or large diameter
ratio. In addition, comparison with the results obtained for $\alpha=0.8$ assuming that
$T_0=T$ shows that the nonequipartition of granular energy has an important influence
on segregation in the absence of gravity. This is consistent with MD-based findings of
Galvin {\em et al.} \cite{GDH05}. Moreover, the results obtained from Eq.\ (\ref{11})
show that the form of the phase diagrams depend significantly on the value of the
reduced gravity $|g^*|$ (namely, reverse buoyancy relative to the effect of Soret
diffusion). Thus, in general for given values of $m_0/m$, $\sigma_0/\sigma$,
$\alpha_0$, $\alpha$ and $\phi$, a transition between both states BNE/RBNE is possible
by changing $|g^*|$. To illustrate the influence of the reduced gravity, a phase
diagram is plotted in Fig.\ \ref{fig3} when $|g^*|=1$ (gravity comparable to the
thermal gradient) for different values of the volume fraction $\phi$. In contrast to
Fig.\ \ref{fig2}, it is apparent that the RBNE regime appears essentially now for both
large mass ratio and/or small size ratio. With respect to the dependence on density,
Fig.\ \ref{fig3} shows that in general the role played by density is quite important
since the regime of RBNE decreases significantly with increasing density. Following
Trujillo {\em et al.} \cite{TAH03}, since the effect of shaking strength of vibration
on the phase diagram BNE/RBNE can be tied to the effect of varying the volume fraction
$\phi$, it is apparent from Fig.\ \ref{fig3} that the possibility of RBNE will increase
with increasing shaking strength. This feature agrees with the experimental findings of
Breu {\em et al.} \cite{BEKR03} since their results show similar behavior with the
external excitation. The form of the phase diagram in the limit $|g^*|\to \infty$ is
shown in Fig.\ \ref{fig4} for $\phi=0.5$ and two values of the coefficient of
restitution $\alpha=1$ and 0.9. Our results indicate that, in contrast to the case of
Fig.\ \ref{fig2}, the main effect of inelasticity is to reduce the size of RBNE region,
which qualitatively agrees again with experiments \cite{BEKR03}. On the other hand, our
predictions disagree with the theoretical results derived by Trujillo {\em et al.}
\cite{TAH03} since the latter found that the mass ratio is a two-valued function of the
size ratio and so the main effect of dissipation is to introduce a threshold value of
the size ratio above which there is no RBNE. The results also indicate (not shown in
Fig.\ \ref{fig4}) that nonequipartition has a weaker influence on segregation when
$|g^*|\to \infty$ than in the opposite limit ($|g^*|=0$). This behavior qualitatively
agrees with the experiments carried out by Schr\"oter {\em et al.} \cite{SUKSS06} for
vibrated mixtures since they find energy nonequipartition to have no discernible
influence on their results.

In summary, a kinetic theory based on a solution of the inelastic Enskog equation has
been used to analyze thermal (Soret) diffusion effects on segregation for an intruder
in a driven moderately dense granular gas under gravity.  The present study goes beyond
the weak dissipation limit, takes into account the influence of both thermal gradient
and gravity on thermal diffusion and applies for moderate densities. Although the
theory is consistent with previous numerical and experimental results, a more
quantitative comparison with the latter would be desirable. As a first test, kinetic
theory predictions in the Boltzmann limit \cite{G06} compare well with MD simulations
of agitated dilute mixtures \cite{BRM05}. Given that the results derived here extends
the description made in Ref.\ \cite{G06} to moderate densities, it can be reasonably
expected that such a good agreement is also kept at finite densities. In this context,
it is hoped that this paper stimulates the performance of such simulations. On the
other hand, it must be stressed that the present work only deals with the tracer or
intruder limit. This precludes the possibility of comparing our theory with the results
reported by Schr\"oter {\em et al.} \cite{SUKSS06} in agitated mixtures constituted by
particles of the same density and equal total volumes of large and small particles.
When convection is practically suppressed, they studied the influence of dissipation on
Soret diffusion. I plan to extend the present theory to finite concentration to carry
out a comparison with the above computer simulation results \cite{SUKSS06} in the near
future.

I am grateful to C. M. Hrenya, J. W. Dufty and A. Santos for useful comments on an
early version of this paper. This work has been supported by the Ministerio de
Educaci\'on y Ciencia (Spain) through grant No. FIS2007-60977, partially financed by
FEDER funds and by the Junta de Extremadura (Spain) through Grant No. GRU08069.


\begin{thebibliography} {99}

\bibitem{K04}A. Kudrolli, Rep. Prog. Phys. {\bf 67}, 209 (2004).

\bibitem{DRC93}J. Duran, J. Rajchenbach, and E. Cl\'ement, Phys. Rev. Lett. {\bf 70},
2431 (1993).

\bibitem{HQL01}D. C. Hong, P. V. Quinn and S. Luding, Phys. Rev. Lett. {\bf 86},
3423 (2001).

\bibitem{BEKR03}A. P. J. Breu, H. M. Ensner, C. A. Kruelle and
I. Rehberg, Phys. Rev. Lett. {\bf 90}, 014302 (2003).

\bibitem{SUKSS06}M. Schr\"oter, S. Ulrich, J. Kreft, S. B. Swift
and H. L. Swinney, Phys. Rev. E {\bf 74}, 011307 (2006).

\bibitem{JY02}J. T. Jenkins and D. Yoon, Phys. Rev. Lett. {\bf 88}, 194301 (2002).


\bibitem{TAH03}L. Trujillo, M. Alam  and H. J. Herrmann, Europhys.
Lett. {\bf 64}, 190 (2003); M. Alam, L. Trujillo and H. J. Herrmann, J. Stat. Phys.
{\bf 124}, 587 (2006).


\bibitem{BRM05}J. J. Brey, M. J. Ruiz-Montero and F. Moreno, Phys. Rev. Lett. {\bf 95}, 098001 (2005).


\bibitem{G06}V. Garz\'o, Europhys. Lett. {\bf 75}, 521 (2006).


\bibitem{SGNT06}D. Serero, I. Goldhirsch, S. H. Noskowicz, and M.-L. Tan, J. Fluid
Mech. {\bf 554}, 237 (2006).


\bibitem{GDH07}V. Garz\'o, J. W. Dufty  and C. M. Hrenya, Phys. Rev. E {\bf 76}, 031303 (2007);
V. Garz\'o, C. M. Hrenya and J. W. Dufty, Phys. Rev. E {\bf 76}, 031304 (2007).


\bibitem{GDH05}J. E. Galvin, S. R. Dahl and C. M. Hrenya, J. Fluid Mech. {\bf 528}, 207 (2005).


\bibitem{DHGD02}S. R. Dahl, C. M. Hrenya, V. Garz\'o and J. W. Dufty, Phys. Rev. E {\bf 66}, 041301 (2002).

\bibitem{FM02}R. Wildman and D. Parker, Phys. Rev. Lett. {\bf 88}, 064301 (2002);
K. Feitosa and N. Menon, Phys. Rev. Lett. {\bf 88}, 198301 (2002).


\bibitem{SBKR05}T. Schautz, R. Brito,
C. A. Kruelle and I. Rehberg, Phys. Rev. Lett. {\bf 95}, 028001 (2005).

\bibitem{MCSL71}N. F. Carnahan and K. E. Starling, J. Chem. Phys. {\bf
51}, 635 (1969); T. Boublik, J. Chem. Phys. {\bf 53}, 471 (1970); E. W. Grundke and D.
Henderson, Mol. Phys. {\bf 24}, 269 (1972).

\bibitem{RG73}T. M. Reed and K. E. Gubbins, {\em Applied Statistical Mechanics} (McGraw-Hill, New York, 1973),
Chap. 6.

\end{thebibliography}
\end{document}